\documentclass[twocolumn]{aastex631}

\usepackage{graphicx}   
\usepackage{amsmath}    
\usepackage{mathrsfs}    
\usepackage{bm}
\usepackage{amssymb}    
\usepackage{diagbox}    
\usepackage{subfigure}
\usepackage{enumerate}
\usepackage{float}
\usepackage{ulem}
\usepackage{url}
\shorttitle{CE ejection from DWDs}
\shortauthors{Y. Zhang et al.}

\begin{document}

\title{Constraints on Common Envelope Ejection from Double Helium White Dwarfs}

\correspondingauthor{Zhenwei Li}
\email{lizw@ynao.ac.cn}

\author{Yangyang Zhang}
\affiliation{Zhoukou Normal University, East Wenchang Street, Chuanhui District, Zhoukou, 466001, People's Republic of China}

\author[0000-0002-1421-4427]{Zhenwei Li}
\affiliation{Yunnan Observatories, Chinese Academy of Sciences, Kunming, 650216, People's Republic of China}
\affiliation{Key Laboratory for the Structure and Evolution of Celestial Objects, Chinese Academy of Science, People's Republic of China}
\affiliation{International Centre of Supernovae, Yunnan Key Laboratory, Kunming, 650216, People's Republic of China}

\author[0000-0001-5284-8001]{Xuefei Chen}
\affiliation{Yunnan Observatories, Chinese Academy of Sciences, Kunming, 650216, People's Republic of China}
\affiliation{Key Laboratory for the Structure and Evolution of Celestial Objects, Chinese Academy of Science, People's Republic of China}
\affiliation{International Centre of Supernovae, Yunnan Key Laboratory, Kunming, 650216, People's Republic of China}

\author[0000-0001-9204-7778]{Zhanwen Han}

\affiliation{Yunnan Observatories, Chinese Academy of Sciences, Kunming, 650216, People's Republic of China}
\affiliation{Key Laboratory for the Structure and Evolution of Celestial Objects, Chinese Academy of Science, People's Republic of China}
\affiliation{International Centre of Supernovae, Yunnan Key Laboratory, Kunming, 650216, People's Republic of China}
\affiliation{University of the Chinese Academy of Science, Yuquan Road 19, Shijingshan Block, 100049, Beijing, People's Republic of China}



\begin{abstract}

Double helium white dwarfs (He WDs) are one type of gravitational wave source and are greatly important in the studies of binary interaction, particularly in the common envelope (CE) ejection physics. Most double He WDs with mass ratios of $q \sim 1$ are formed through a particular channel. In this channel, one He WD is initially produced from a red giant (RG) with a degenerate core via stable Roche lobe overflow, and another He WD is formed from an RG with a degenerate core via CE ejection.
They may have significant implications for the binary evolution processes yet have not received specific studies, especially for the CE phase. This paper adopts a semi-analytic method and a detailed stellar evolution simulation to model the formation of double He WDs.
We find that most double He WDs show mass ratios being slightly greater than 1, and their orbital periods and mass ratios relation are broadly consistent with observations.
There is also a relation between the mass ratios and the progenitors' masses of the He WDs produced via CE ejection for double He WDs with determined WD masses.
Based on this relation, the mass of the He WD progenitor can be inferred from the mass ratio. Then, the CE ejection efficiency can be constrained with the orbital period. 
In addition, we constrain the CE ejection efficiency for two double He WDs, J1005-2249 and WD0957-666. The results show that the CE ejection efficiencies increase with the WD progenitor masses.

\end{abstract}



\section{Introduction}
\label{sec:1}

Double white dwarfs (WDs) play important roles in studying the binary evolution processes, such as stable Roche lobe overflow (RLOF) and common envelope (CE) ejection phase \citep{Rappaport1982, Warner1995}. Double WDs in close orbits are presumed to be SN Ia progenitors and also one type of important gravitational wave source \citep{Postnov2006, LiuZhengwei2023, Amaro-Seoane2023}. Some close double WDs in electromagnetic observations are also detectable sources for future space-based gravitational-wave detectors, such as Laser Interferometer Space Antenna \citep{Amaro-Seoane2017, Kupfer2018, Brown2020, Kupfer2022}. For example, ZTFJ2243+5242 likely consists of two Helium (He) WDs, and LISA can detect this system due to its short orbital period (8.8 minutes)\citep{Burdge2020b}.

Double WDs are supposed to be formed via at least one mass transfer phase. The main formation channels of Double WDs are summarized as follows \citep{Iben1997, Han1998, Nelemans2001b, Li2023, Li2024}. The RLOF+CE channel: the relatively massive star (primary star) in a binary evolves faster and fills its Roche Lobe, undergoes stable RLOF, and leaves a WD. As the system evolves, the secondary star fills its Roche Lobe and enters into the CE phase due to the unstable mass transfer.
If the secondary envelope can be successfully ejected, a double WDs is formed. The CE+CE channel: this process is similar to the RLOF+CE channel, but the primary star evolved to be a WD through the CE ejection. Besides these two main formation channels, there are some other formation channels that form the double WDs. The CE+RLOF channel: the primary star evolves to be a WD through CE ejection. As the secondary evolves to fill its Roche Lobe, the mass transfer is stable, leaving a double WD after the stable mass transfer. This channel mainly produces the double WD with extremely low-mass WD (ELM WD; \citealt{Li2019}). The RL+RL channel: the two mass transfers are stable in this channel. Double WDs formed through this channel usually have wide orbits ($\gtrsim 1 \rm{au}$; \citealt{Korol2022a}). The single CE channel: the binaries have an initial mass ratio close to 1; they enter the CE phase when both stars are on the red giant branch (RGB) or asymptotic giant branch (AGB). The system can also evolve into double WDs after both envelopes of the two stars' ejection. This channel accounts for a very small fraction of double WDs \citep{Han1998}.     

Double WDs formed through the RLOF+CE channel generally have mass ratios of $q \sim 1$ ($q = M_{\rm{WD,2}}/M_{\rm{WD,1}}$), where $M_{\rm{WD,1}}$ means the mass of the first born WD; $M_{\rm{WD,2}}$ is the second born WD mass \citep{Li2023}. 
The reason is introduced as follows: if the massive primary undergoes the stable RLOF on the RGB or AGB with a degenerate core, the core masses correlate with the radii of the giant stars \citep{Refsdal1971, Webbink1983}. Then, assuming that the star radii are equal to the Roche lobe radii, a tight relation exists between the masses of WDs and binary separations (orbital periods) after the stable RLOF. \citep[e.g.,][]{Savonije87, Joss1987, Rappaport1995, Tauris1999, Chen2013}.
As the secondary evolves and fills its Roche lobe with a degenerate core, the binary separation also determines its radius and core mass. The latter is close to the WD mass after the CE ejection.
Then, the two WD masses both have relationships with the binary separation at the termination of the first mass transfer.
So, there is a relation between the two WD masses. \citet{Li2023} show that double WDs have mass ratios $q \sim 1$, which differs from the mass ratios ($q \sim 0.25$) of the double WDs produced from the CE+CE channel. However, they do not give the relations among the progenitor mass of WD, the second-born WD mass and the mass ratio.

Double He WDs with mass ratios of $q \sim 1$ are more likely formed through the stable RLOF + CE channel and their progenitors are red giants (RG) with degenerate cores. Donors with mass $\gtrsim 2.0M_\odot$ would develop a non-degenerate core and contribute to the formation of double He WDs. However, this contribution is supposed to be small ($\lesssim 15\%$; depending on the CE efficiencies, see \citealt{Li2023}) based on the initial mass function.  
Therefore, for the double He WDs formed through this particular channel, we expect to find the relationship between the mass ratio of the double WDs and the progenitor mass of the second He WD. Then, we can constrain the CE efficiency from the progenitor mass of the second He WD and the orbital period. 

\citet{Scherbak2023} studied the CE efficiency with double WDs, and they suggested the CE efficiency is around 1/3. However, they did not constrain the progenitor masses of WDs with the mass ratios. 
This paper aims to investigate the relationship among the mass ratio, orbital period, the progenitor's mass of second-born He WDs and CE ejection efficiency of double He WDs with a semi-analytic method and detailed binary evolution simulations. These methods have been used to study the novae, pre-cataclysmic variable systems, and sdB + He WD binaries \citep[e.g.,][]{Shen09, Nelson2018, Zhang2021b}.
The paper is structured as follows. We use a semi-analytic method to model the double He WDs formation and derive its orbital period and mass ratio relation in Section 2. We follow the formation of double He WDs with a detailed binary evolution code in Section 3.
Then, we obtain the double He WDs' orbital period and mass ratio relation.
The comparison between the theoretical relation and the observational data of double He WDs is addressed in Section 4. For the observed DWDs of WD0957-666 and J1005-2249, we constrain the CE efficiencies and the progenitor masses of the second He WDs with the high precision mass ratios, as shown in Section 5. The summary and discussion are given in Section 6. 

\section{The orbital period - mass ratio relation from a semi-analytic method}
\label{sec:2}

As introduced above, double He WDs with mass ratios of $q\sim 1$ are likely formed through the stable RLOF+CE channel. We first study this issue with a semi-analytic method.

In the stable RLOF+CE channel, the primary evolves to the RG with a degenerate core;
there exists a relation between the He core mass (${M_\mathrm{c}}$) and the star radius ($R$), i.e., ${M_\mathrm{c}}$ - $R$ relation \citep{Refsdal1971, Webbink1983, Joss1987}
, which is given by \citet{Rappaport1995} \footnote{The relationship between WD mass and orbital period obtained from this formula is not highly accurate, especially at a low orbital period (or equivalently a small WD mass) \citep{Tauris1999, Chen2013,Lin2011},  so we get the actual core mass, star radius, and other parameters from the detailed binary evolution model in Section 3 rather than use this formula.}:

\begin{equation}
\label{1}
R =  \frac{{\rm R_{0}}M_{\mathrm {c}}^{4.5}}{1+4M_{\mathrm {c}}^{4}}+0.5 \mathrm{R_{\sun}} \equiv f_{1} (M_{\rm c}),
\end{equation}
where $R$ is the star radius in units of solar radius, and $M_{\rm c}$ is the degenerate core mass in units of solar mass. ${\rm R_{0}}$ equals 5500 $\rm R_{\sun}$ for metallicity of $Z = 0.02$ (Population I) and 3300 $\rm R_{\sun}$ for metallicity of $Z = 0.001$ (Population II). We use the function $f_{1}(M_{\rm c})$ to show this relation for convenience. In the following, we will employ similar symbols $Y \equiv f(X_{1}, X_{2})$ to demonstrate the dependence of parameter $Y$ on others ($X_{1}$, $X_{2}$).

For the Roche lobe radius ($R_{\rm L}$) of the primary, it is determined by the mass ratio ($q$) and the binary separation ($a$) \citep{Peter1983}:
 \begin{equation}
 \label{2}
R_{\mathrm{L}}/a \approx \frac{0.49q^{2/3}}{0.6q^{2/3}+\mathrm{ln}(1+q^{1/3})} \equiv f_{2} (q).
\end{equation}

During the stable RLOF phase, the radius of the primary star is approximately equal to its Roche lobe radius, i.e., $R \approx R_{\rm L}$. There is a relation between the binary separation, $a$, and the degenerate core mass, $M_{\rm c}$:
\begin{equation}
\begin{split}
\label{3}
a & = R_{\rm {L}} / f_{2} (q) \approx R / f_{2} (q) = f_{1} (M_{\rm {c}}) / f_{2} (q)\\
                          & \equiv  f_{3} (M_{\rm c}, q).
\end{split}
\end{equation}

After the mass transfer phase, the core mass of the primary ($M_{\rm c,1}$) is approximately equal to the final WD mass ($M_{\rm WD,1}$).
Then, we can get the binary separation after the first stable RLOF ($a_{\rm f,RLOF}$) with Eq.~\ref{3}:
\begin{equation}
\begin{split}
\label{4}
a_{\rm {f, RLOF}} & = f_{3} (M_{\rm {c, 1}}, q_{1}) \approx  f_{3} (M_{\rm WD,1}, M_{\rm {\rm {WD,1}}}/M_{\rm 2}) 
\end{split}
\end{equation}
where $M_{2}$ is the mass of secondary (the progenitor's mass of the second WD) at the end of stable RLOF.

We assume the binary separation will not change until the secondary fills its Roche lobe, i.e., the binary separation after the stable RLOF is equal to the binary separation at the beginning of the CE phase ($a_{\rm i,CE}$).
Although the binary separation can change with gravitation-wave radiation (GWR), magnetic braking (MB), and tides, their effect is weak due to the relatively large binary separation. At the onset of the CE phase, if the secondary also has a degenerate core, the star core mass and radius should follow the $M_{\rm c}$-$R$ relation as well. Then, by means of Eq.~\ref{3}, we can get the relation between the $a_{\rm i,CE}$ and the core mass of the secondary ($M_{\rm {c,2}}$) at the beginning of the CE phase:
\begin{equation}
\begin{split}
\label{5}
a_{\rm {f, RLOF}} = a_{\rm i, CE} = f_{3} (M_{\rm {c,2}}, q_{2})\\
\approx f_{3} (M_{\rm {WD, 2}}, M_{\rm {2}}/M_{\rm {WD,1}}),
\end{split}
\end{equation}
where $M_{\rm{WD,2}}$ is the second WD's mass. Generally, it should be approximately equal to the $M_{\rm {c,2}}$.

From Eq.~\ref{4} and ~\ref{5}, we find a relation between the $M_{\rm {WD,1}}$, $M_{\rm {WD,2}}$ and $ M_{\rm {2}}$, i.e.,
\begin{equation}
\label{6}
M_{\rm {WD,1}} \equiv f_{4} (M_{\rm {WD,2}}, M_{\rm {2}}).
\end{equation}

For the CE ejection, we adopt the energy budget prescription to describe this process \citep{Webbink1984}:
 \begin{equation}
 \label{7}
E_{\mathrm{bind}} = \alpha_{\mathrm {ce}}(\frac{GM_{\mathrm{WD,1}}M_{\mathrm{WD,2}}}{2a_{\mathrm{f, CE}}} - \frac{GM_{\mathrm{WD,1}}M_{2}}{2a_{\mathrm{i, CE}}}),
\end{equation}
where $G$ is the gravitational constant; $\alpha_{\mathrm {ce}}$ is the efficiency of the released orbital energy. $E_{\mathrm {bind}}$ is the binding energy of the H-rich envelope. It includes the gravitational energy ($E_{\mathrm {gr}}$) and thermal energy ($E_{\mathrm {th}}$) \citep{Han2002, Han2003}:
\begin{equation}
 \label{8}
 E_{\mathrm {bind}}  =  -(E_{\mathrm {gr}} + \alpha_{\mathrm {th}} E_{\mathrm {th}}).
\end{equation}
Here, $\alpha_{\mathrm {th}}$ is the efficiency of thermal energy. For a fixed $\alpha_{\mathrm {th}}$, the binding energy is related to the mass of the secondary and its core mass (i.e., the second WD mass):
\begin{equation}
 \label{9}
 E_{\mathrm {bind}}  \equiv  f_{5} (M_{2}, M_{\rm {c, 2}},\alpha_{\mathrm {th}}) =  f_{5} (M_{2}, M_{\rm {WD,2}},\alpha_{\mathrm {th}}) .
\end{equation}
We present this relation in Appendix~\ref{App} for $\alpha_{\mathrm {th}}$ of 1. Additionally, the binding energy can be interpolated for various  $\alpha_{\mathrm {th}}$.

Then, after the CE ejection, we can get the binary separation with Eq.~\ref{5},~\ref{6},~\ref{7} and~\ref{9}:
\begin{equation}
\begin{split}
\label{10}
a_{\mathrm {f, CE}} & =  \frac{GM_{\mathrm{WD,1}}M_{\mathrm{WD,2}}}{2E_{\mathrm{bind}}/\alpha_{\mathrm {ce}}+GM_{\mathrm {WD,1}}M_{2}/{a_{\mathrm {i, CE}}}} \\
  & \equiv f_{6} (M_{\rm {WD,1}}, M_{\rm {2}},\alpha_{\mathrm {th}},\alpha_{\mathrm {ce}}).
\end{split}
\end{equation}

We can also obtain the orbital period with Kepler's third law after the CE phase:
\begin{equation}
 \label{11}
\begin{split}
 P_{\rm {orb}} & = \sqrt{\frac{4\pi^{2} a_{\mathrm {f, CE}}^{3}}{G(M_{\mathrm{WD,1}}+M_{\mathrm{\rm WD,2}})} } 
 \equiv f_{7} (M_{\rm {WD,1}}, M_{\rm {2}},\alpha_{\mathrm {th}},\alpha_{\mathrm {ce}})\\ 
 & \equiv f_{8} (M_{\rm {WD,1}}, q_{\rm{f}},\alpha_{\mathrm {th}},\alpha_{\mathrm {ce}}). 
\end{split}
\end{equation}

\begin{figure}
   \centering
   \includegraphics[width=\columnwidth]{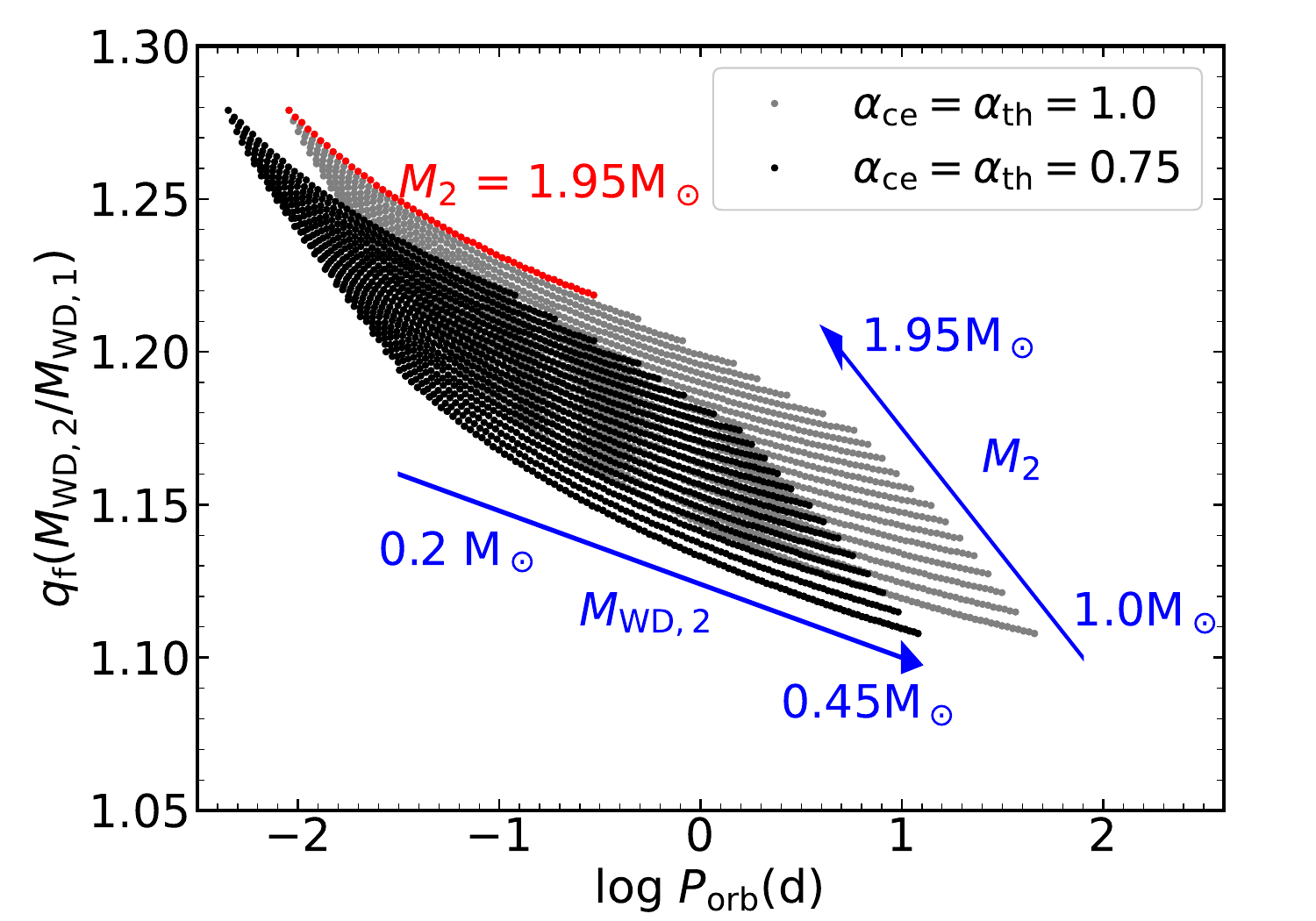}
      \caption{The orbital period and mass ratio relation of double He WDs is obtained by the semi-analytic method. The grey and black dots represent models with $\alpha_{\rm ce} = \alpha_{\rm th} = 0.75$ and $\alpha_{\rm ce} = \alpha_{\rm th} = 1.0$, respectively. The progenitor mass of the second WD ranges from 0.8 $\rm M_{\odot}$ to 1.95 $\rm M_{\odot}$ and increases from the lower right to the upper left. The mass of the second He WDs ranges from $\sim0.2$ $\rm M_{\odot}$ to $\sim0.45$ $\rm M_{\odot}$ and increases from the upper left to the lower right. To better understand the orbital period and mass ratio relation for different $M_{\rm{2}}$, we choose the relation for $M_{\rm{2}}=$ 1.95 $\rm M_{\odot}$ with the red color.}
         \label{P-qf-semi}
\end{figure}

According to this equation, there is a relation between the $P_{\rm {orb}}$, $q_{\rm{f}}$ ($M_{\rm {WD,2}}$/$M_{\rm {WD,1}}$), $M_{\rm {WD,1}}$ (or $M_{\rm {WD,2}}$), $\alpha_{\mathrm {ce}}$, and $\alpha_{\mathrm {th}}$. 
With Eq.~\ref{6}, the $M_{\rm {WD,1}}$ are related to the $M_{2}$ and $M_{\rm {WD,2}}$. The relations between $P_{\rm {orb}}$ and $q_{\rm{f}}$ with different CE efficiencies and $M_{2}$ are shown in Fig.~\ref{P-qf-semi}. 

Fig.~\ref{P-qf-semi} shows that the mass ratios of these systems are slightly larger than 1, i.e., $q_{\rm {f}} \sim 1.1-1.3$. The reason for the relatively larger masses of the second-born WDs can be understood as follows. The binary separation after the first RLOF determines not only the first-born WD's mass but also the secondary Roche lobe radius (see Eq.~\ref{4} and~\ref{5}). Since the secondary is no doubt more massive than the WD produced at the end of RLOF (i.e., $M_2/M_{\rm {WD,1}} \sim 2-10$), the secondary's Roche lobe radius ($R_{\rm{L,2}}$) is larger than the Roche lobe ($R_{\rm{L,1}}$) (or the radius) of the primary (i.e., $R_{\rm{L,2}}/R_{\rm{L,1}} \sim 1.3-2.8$ as obtained from Eq.~\ref{2}). Consequently, the core mass of the secondary (according to the core-radius relation of red giants, i.e., Eq.~\ref{1}) when it fills its Roche lobe is more massive than the first-produced WD, resulting in a more massive WD after the ejection of the envelope. The secondary mass (or the mass ratio of the binary) after the first RLOF gives a further constraint on its Roche lobe before the CE, which means that a larger $M_2$ results in a larger Roche lobe, then a more massive WD ($M_{\rm {WD,2}}$) and a relatively larger $q_{\rm {f}}$ consequently. This can also be seen in Fig.~\ref{P-qf-semi}. For a given $M_2$, the binding energy of the H-rich envelope decreases with the increase of $M_{\rm {WD,2}}$, resulting in the increase of the orbital period.

From Eq.~\ref{11}, if the WD masses are derived from the light curves or other methods, and the mass ratio inferred from the two WD masses or radial velocity and so on, the $\alpha_{\mathrm {th}}$ and $\alpha_{\mathrm {ce}}$ can be derived from the orbital period. In order to get a more precise result, we model the formation of double He WDs with detailed binary evolution code in the next section.

\section{The orbital period and mass ratio relation from detailed binary evolution calculation} 

In order to better understand the relation derived above, we adopt the stellar evolution code \texttt{Modules for Experiments in Stellar Astrophysics} (\texttt{mesa}, version 10398; \citealt{Paxton2011, Paxton2013, Paxton2015, Paxton2018, Paxton2019}) to model the formation of double He WDs from main sequence (MS) binaries in this section.

\subsection{Binary evolution models}

The main inputs in \textsc{mesa} code are summarized as follows. The chemical compositions, mixing length, overshooting, stellar wind, and opacity table are the same as in Appendix~\ref{App}, which we adopted for calculating the binding energy in Section \ref{sec:2}.
In the binary model, the initial primary star masses are equal to 1.26, 1.6, 1.8, 1.95 $\rm M_{\odot}$ at $Z = 0.02$ and 1.0, 1.26, 1.6, 1.85 $\rm M_{\odot}$ at $Z = 0.001$. The secondary masses range from 1.0 to 1.9 $\rm M_{\odot}$ for $Z = 0.02$ and range from 0.8 to 1.8 $\rm M_{\odot}$ for $Z = 0.001$ in steps of $\Delta{M} = 0.1 {\rm M_{\odot}}$, in which the secondary can evolve to the RG. The initial orbital periods range from 0.6 to 150 days in logarithmic steps of $0.1$, in which range the binaries systems can form to be the double He WDs.

We adopt the "Kolb" scheme in calculating the mass transfer rate \citep{Kolb1990}. This scheme takes into account the stellar atmosphere, and then the mass transfer can occur when the radii of the donors are smaller than its Roche lobe radii.

As for the angular momentum loss, we consider the gravitational-wave radiation, magnetic braking, and mass loss. The angular momentum loss due to GWR followed the \citet{Landau1971}:
\begin{equation}
 {\dot J_\mathrm{GR}}  = - \frac{32}{5} \frac{G^{7/2}}{c^5} \frac{M_\mathrm{1}^2 M_\mathrm{2}^2 (M_\mathrm{1}+M_\mathrm{2})^{1/2}}{a^{7/2}}
\end{equation}
where $c$ is the speed of light in vacuum; $M_\mathrm{1}$ and $M_\mathrm{2}$ are the masses of the primary and secondary, respectively. ${a}$ is the binary separation.

The angular momentum loss due to MB follows the equation by assuming index $ \gamma = 3$ \citep{Rappaport1983}:
\begin{equation}
 {\dot J_\mathrm{MB}}  = -3.8 \times 10^{-30} M_\mathrm{1} R_\mathrm{1}^{3} \omega^{3} \rm dyn \cdot cm
 \end{equation}
where $M_\mathrm{1}$ and $R_\mathrm{1}$ are the mass and radius of the star, respectively. $\omega$ is the star's rotational angular velocity, and we assume the system is tidally synchronized, i.e., $\omega$ is also equal to the orbital angular velocity. We consider the MB for a star with a convective envelope and switch off the MB when the mass of the surface convective region is less than 1 percent of the total mass \citep{Chen2017}. The MB is considered for both the primary and secondary in our calculations.

The mass transfer efficiency ($\beta$) is still an open issue. Due to the uncertainty of the specific value of mass transfer efficiency, for convenience, we assume the accretion efficiency is equal to 0 in the first mass transfer. Besides, the unprocessed material will carry away the specific angular momentum of the accretor. This assumption will not influence the ${M_\mathrm{c}}$ - $R$ relation (also Eq.~(\ref{6}) and (\ref{11})), so it exerts a minor effect on our results. We discuss the influence of mass transfer efficiency in Appendix ~\ref{App2}. If the first mass transfer rate is larger than $10^{-4}\;\mathrm{M_{\sun} yr^{-1}}$, we consider the mass transfer to be unstable, leading to the  
occurrence of the CE phase \citep{Chen2017}.

At the onset of the second mass transfer, the mass ratios of the binaries are large (1.7 - 6.3), so we assume all the binaries will enter into the CE phase \citep{Han2002, Chen2008}. Then, we compute the binding energy for a given $\alpha_{\rm th}$ with Eq.~\ref{8}. Using Eq.~\ref{7},~\ref{10} and~\ref{11}, we can obtain the properties of double He WDs after CE ejection for a given $\alpha_{\rm ce}$.

We end our simulations once the second mass transfer happens or the evolution time exceeds 15 Gyr. Then, we pick out the binaries that can evolve into double He WDs. All of the \textsc{mesa} input files and the data required to reproduce our results are available at \dataset[doi:10.5281/zenodo.11125846]{https://doi.org/10.5281/zenodo.11125846}.

\subsection{Binary evolution results}
\begin{figure}
    \centering
    \includegraphics[width=\columnwidth]{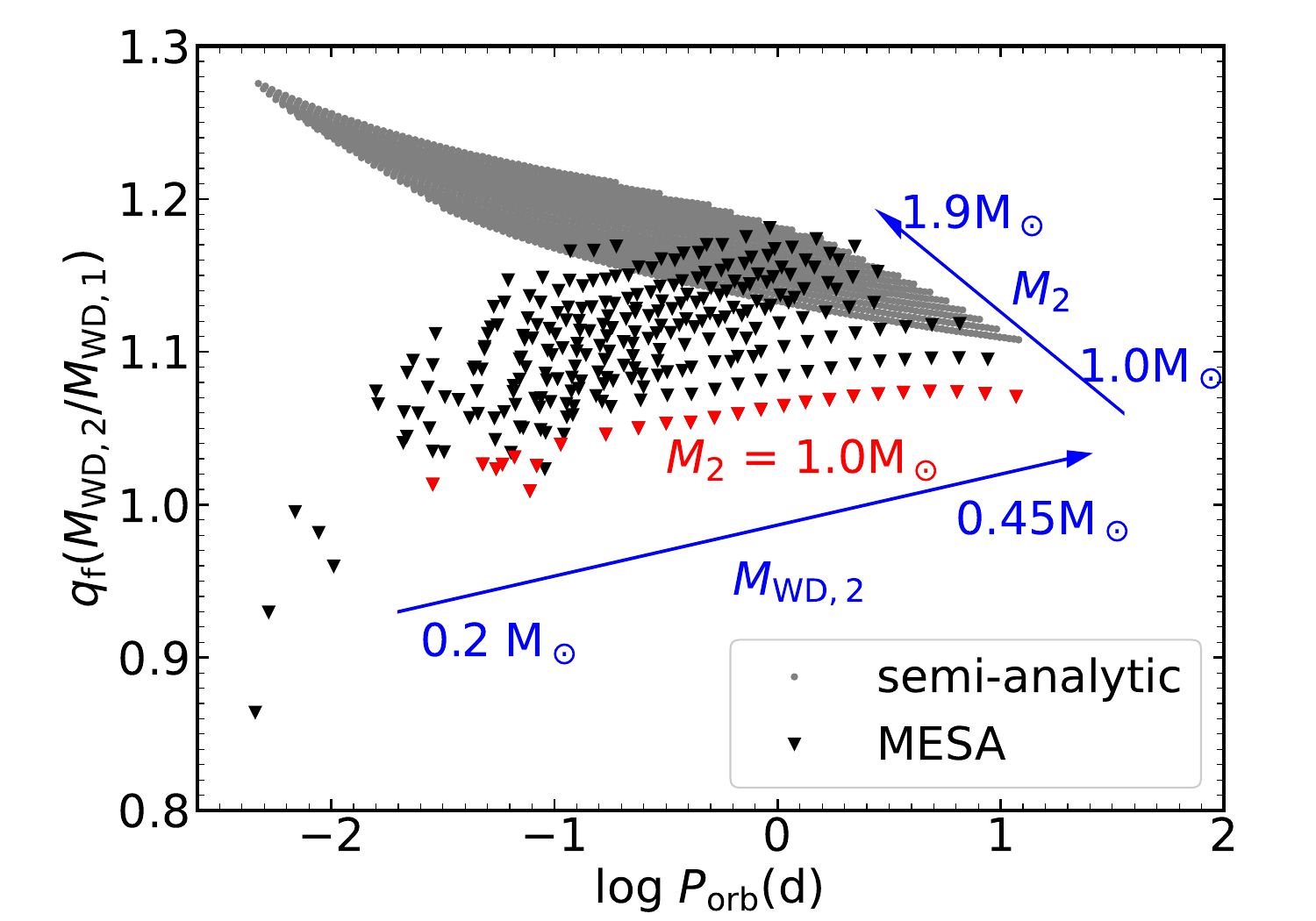}
    \caption{Comparison of orbital period and mass ratio relations derived from the detailed binary evolution (triangles) and semi-analytic (dots) with Z = 0.02. In these models, $\alpha_{\rm ce} = \alpha_{\rm th} = 0.75$ are adopted. Similar to the Fig.~\ref{P-qf-semi}, we choose the orbital period and mass ratio relation for $M_{\rm{2}}=$ 1.0 $\rm M_{\odot}$ with the red color.}
    \label{fig:P-q-s-m}
\end{figure}

From the above binary computations, we show the relation between the orbital period and mass ratio for the double He WDs in Fig.~\ref{fig:P-q-s-m}. The relationship derived from the semi-analytic method is also included for comparison.
This plot shows that the mass ratios obtained by the detailed binary evolution are approximately equal to 1, and the mass of the second WD is larger than the first WD in most cases. It is consistent with the result from the semi-analytic method. We notice that the mass ratios from these two methods are different, especially for the short orbital period. The reason is as follows: Firstly, in the \textsc{mesa} calculation, a thin H envelope around the He core after the first mass transfer makes the first WD mass larger than its core mass, while the mass of the first WD is precisely equal to the core mass in the semi-analytic method.
Second, in the detailed stellar evolution calculation, the core mass and radius slightly deviate from the $M_{\rm c}-R$ relation given in Equation~\ref{1}. Especially for the short-orbital-period double He WDs, the primary star begins the stable RLOF at the end of the main sequence (MS) or the Hertzsprung Gap (HG); its core mass and radius fit the $M_{\rm c}-R$ relation not well (a small radius corresponds to a larger core mass and then a larger WD). These two reasons make the first WD from \textsc{mesa} larger than the first WD from the semi-analytic method. Then, the mass ratio in the \textsc{mesa} calculation is smaller than the result from the semi-analytic method.
In addition, during the first mass transfer or at the beginning of the CE phase, the donor's radius is assumed to be equal to its Roche lobe radius in the semi-analytic method, which is different from the \textsc{mesa} calculation (i.e., the donor's radius can be smaller than its Roche lobe radius). Moreover, the binding energies calculated through interpolation might be slightly overestimated in the semi-analytic method.

\section{Comparison with Observations}
\label{sec:4}
\begin{figure}
    \centering
    \includegraphics[width=\columnwidth]{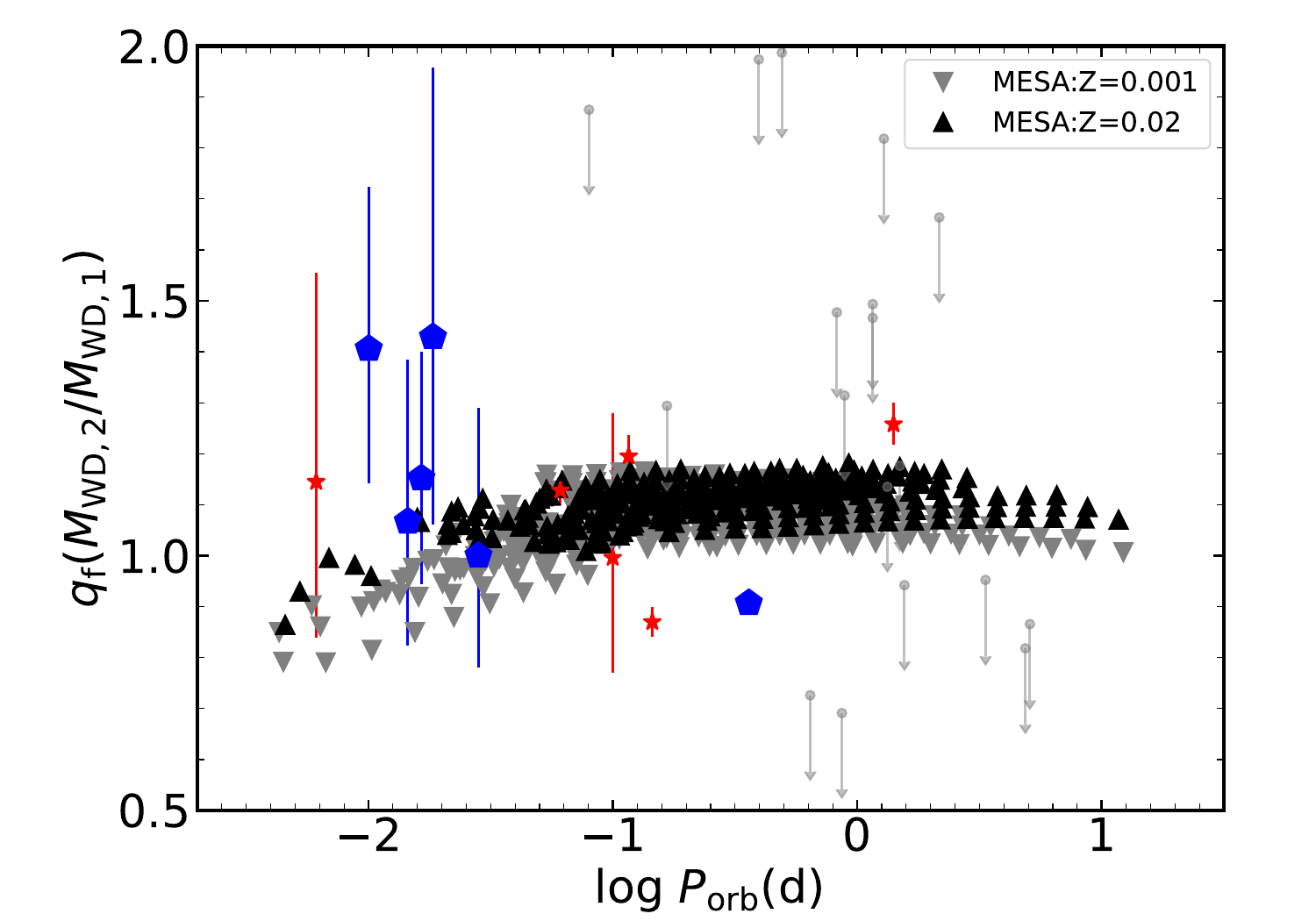}
    \caption{Comparison of orbital period and mass ratio relations for double He WDs from our calculations with observations. The gray dots with arrows represent possible double He WDs with only minimum companion mass ($M_{\rm WD,1}$) determined \citep{Kruckow2021, Li2023}; we only show the systems with mass ratios smaller than 2.
    The systems with red stars are for the double He WDs and the mass ratios constrained by the radial velocities (RV) \citep{Burdge2020a, Maxted2002b, Hallakoun2016, Bours2014, Brown2011a, Maxted1999}. The pentagons represent the mass ratios constrained by the two WD masses \citep{Burdge2020b, Coughlin2020, Brown2013, Napiwotzki2020}. Black and gray triangles indicate the results of the \textsc{mesa} models, corresponding to metallicities of $Z = 0.02$ and $Z = 0.001$, respectively. In these models, we adopt the CE efficiencies $\alpha_{\rm ce} = \alpha_{\rm th} = 0.75$. 
    }
    \label{fig:com_obs}
\end{figure}
\citet{Kruckow2021} have collected a sample of potential post-common envelope binaries (see also table C.1 in \citealt{Li2023}). It is hard to distinguish He WDs from these samples. In our calculation, if the He core mass is larger than $0.47\;{\rm M_{\odot}}$ for Population II ($Z = 0.001$) or $0.45\;{\rm M_{\odot}}$ for Population I ($Z = 0.02$), the core can be ignited and cannot form He WD. So, we select the double WDs with either WD mass smaller than $0.47\;{\rm M_{\odot}}$ as double He WDs. It is important to note that some of the WDs within this sample may not be pure He WDs, i.e., hybrid HeCO WDs with carbon-oxygen (CO) cores. Following \citet{Li2023}, we performed the binary population synthesis and found that the fraction of hybrid WDs relative to WDs with masses $\lesssim 0.45M_\odot$ is $\lesssim 30\%$ (the fraction of hybrid WDs may be affected by the initial input parameters in the binary population synthesis simulations). Therefore, WDs in these samples are more likely He WDs \footnote{Although the majority of WDs are not likely to be hybrid WDs, the hybrid WDs departure from the radius-core mass relationship of Eq.~\ref{1} can be quite radical. In this case, the mass ratios of binaries with hybrid WDs are different from those containing He WDs.}. 

There are six double He WDs given the radial velocities in the sample of \citet{Kruckow2021}, so we constrain the mass ratios ($q = M_{\rm WD,2}/M_{\rm WD,1} = K_{1}/K{2}$, where $K_{1}$,$K_{2}$ are the RV semi-amplitudes of the first WD and the second WD, respectively) with the radial velocities. There are six double He WDs that only have the WD mass, and then we get the range of mass ratios with the two WD masses. In the other samples, only the minimum masses are constrained for the first WD, so we only give the maximum mass ratio for these double WDs.
 
We compare the orbital period and mass ratio relation of double He WDs with these observational data in Fig.~\ref{fig:com_obs}.
There are seven double He WDs with a maximum mass ratio larger than 2; one even has the maximum mass ratio $q_{\rm{f}} \sim 11$. These systems have an uncertainty of the mass ratios and little influence on our result, so we only show the double WDs with a mass ratio smaller than 2. In the plot, we see the most observed double He WDs with mass ratios $q_{\rm{f}} \sim 1$ and the second WD mass is larger than the first WD. The mass ratios constrained by the WD mass have a larger error bar due to the uncertainty of the WD masses. The systems have the RV semi-amplitudes; some have a well-constrained mass ratio due to the high-precision RV semi-amplitudes that are obtained from the spectrum. For example, the J1005-2249 (CSS 41177) and WD0957-666 are the two special systems we will use to constrain the CE efficiencies in the next section. The plot shows some observed double He WD with a mass ratio smaller than 1. This system has a wide binary separation, so the binary model from our calculation cannot cover these observations with the larger CE efficiencies. For this kind of double He WDs, some may be from another formation channel, such as the CE+CE channel or the single CE channel, or their progenitor's masses are larger than $2.0\;\rm M_{\odot}$. Moreover, some of them may be CO WDs.    

By comparing our results with observations, we also see that our calculation broadly agrees with observation. In our calculation, the orbital period - mass ratio relation is similar for Population I and Population II.

\section{Studying the CE efficiencies with the mass ratio}
From Eq.~\ref{6}, if we know one of the WD masses and the mass ratio for a double He WDs, we can know the progenitor mass of the second WD. Then, we can constrain the CE efficiencies with the orbital period. In this section, we constrain the CE efficiencies with the high-precision mass ratios for the J1005-2249 and WD0957-666. The WD masses of the two systems are smaller than 0.4 $\rm M_{\odot}$; we assume all of them are He WD and the progenitor masses of them are smaller than $2.0\rm M_{\odot}$, i.e., the progenitors have degenerate cores on the red giant (according to the population synthesis results of \citet{Li2023}, this probability of He WDs' progenitors with masses less than $2.0M_\odot$ is $\geq 80\%$). 
The binding energy will change with the metallicity, which affects the CE efficiencies. Since DWDs are more likely found in the field, the metallicity is set to be $Z=0.02$, similar to \citet{Scherbak2023}.

\subsection{J1005-2249}

\begin{figure}
    \centering
    \includegraphics[width=\columnwidth]{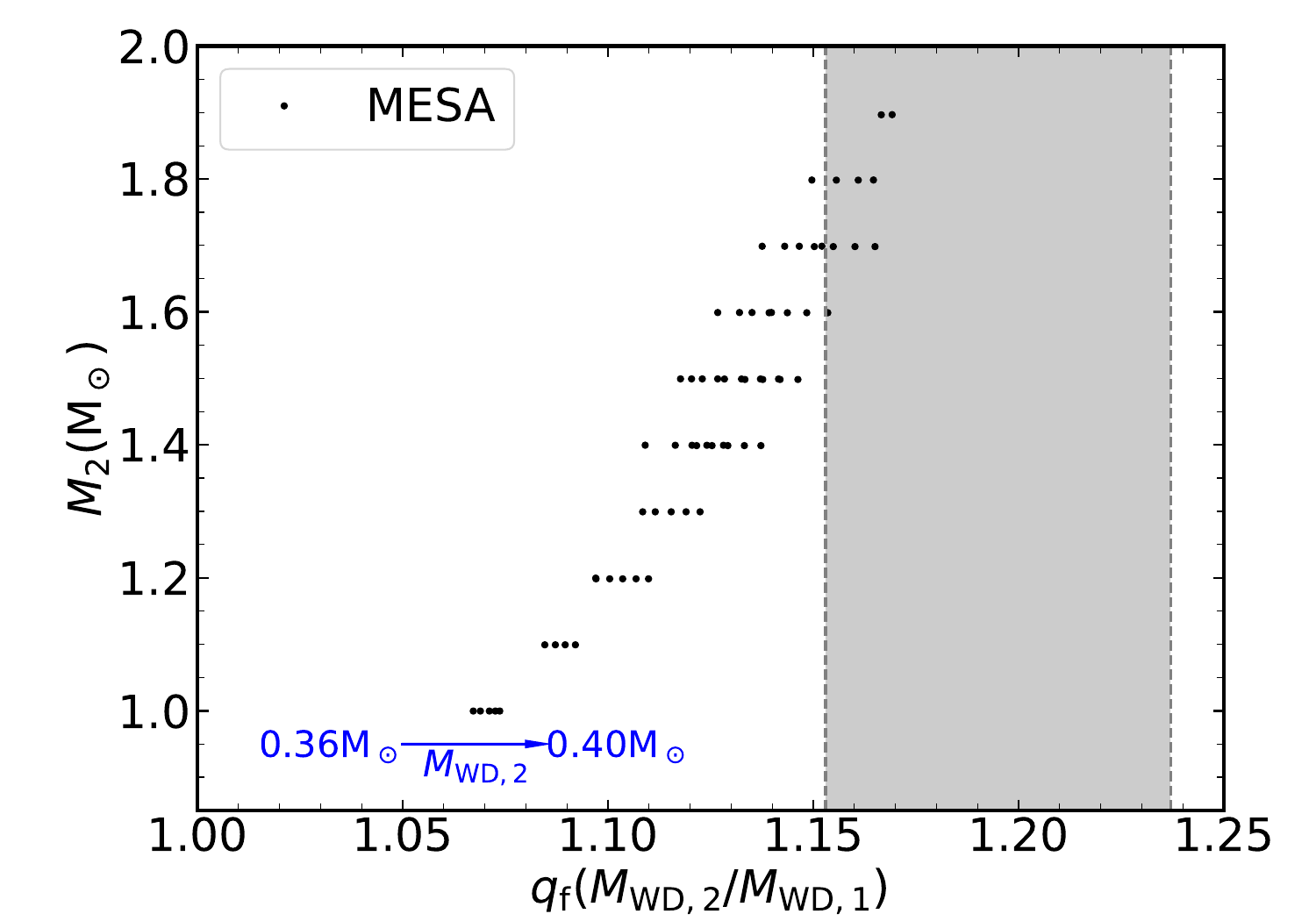}
    \caption{The $q_{\rm f}$ and $M_{\rm 2}$ for J1005-2249 compared to the models from \texttt{mesa}. The gray dots represent the models from \texttt{mesa} with $Z = 0.02$. The gray shade represents $q_{\rm f}$ and the progenitor mass ($M_{\rm 2}$) for the J1005-2249.}
    \label{fig:q-m-J1005}
\end{figure}

The J1005-2249 is a double-lined spectroscopic double WDs. \citet{Bours2014} give the masses for both WD by combined modeling of the light curves and radial velocities. The first WD mass is $M_{\rm WD,1} = 0.32\pm 0.01 \rm M_{\odot}$, and the second is $M_{\rm WD,2} = 0.38 \pm 0.02 \rm M_{\odot}$. The radial velocity amplitudes are $K_{1} = 210.4 \pm 6.1 {\rm km s^{-1}}$ and $K_{2} = 176.1 \pm 1.1 {\rm km s^{-1}}$ for the first and second WD, respectively. Then, the mass ratio of J1005-2249 is $q = M_{\rm WD,2}/M_{\rm WD,1}=K_{1}/K{2}= 1.195_{-0.031}^{+0.042}$. In Fig.~\ref{fig:q-m-J1005}, we show the $q_{\rm f}$ and $M_{2}$ relation by adopting the $M_{\rm WD,2} = 0.38 \pm 0.02 \rm M_{\odot}$ based on Eq.~\ref{6}. We find that the progenitor mass ($M_{2}$) of J1005-2249 range from 1.6 $\rm M_{\odot}$ to 2.0 $\rm M_{\odot}$. 

\begin{figure}
    \centering
    \includegraphics[width=\columnwidth]{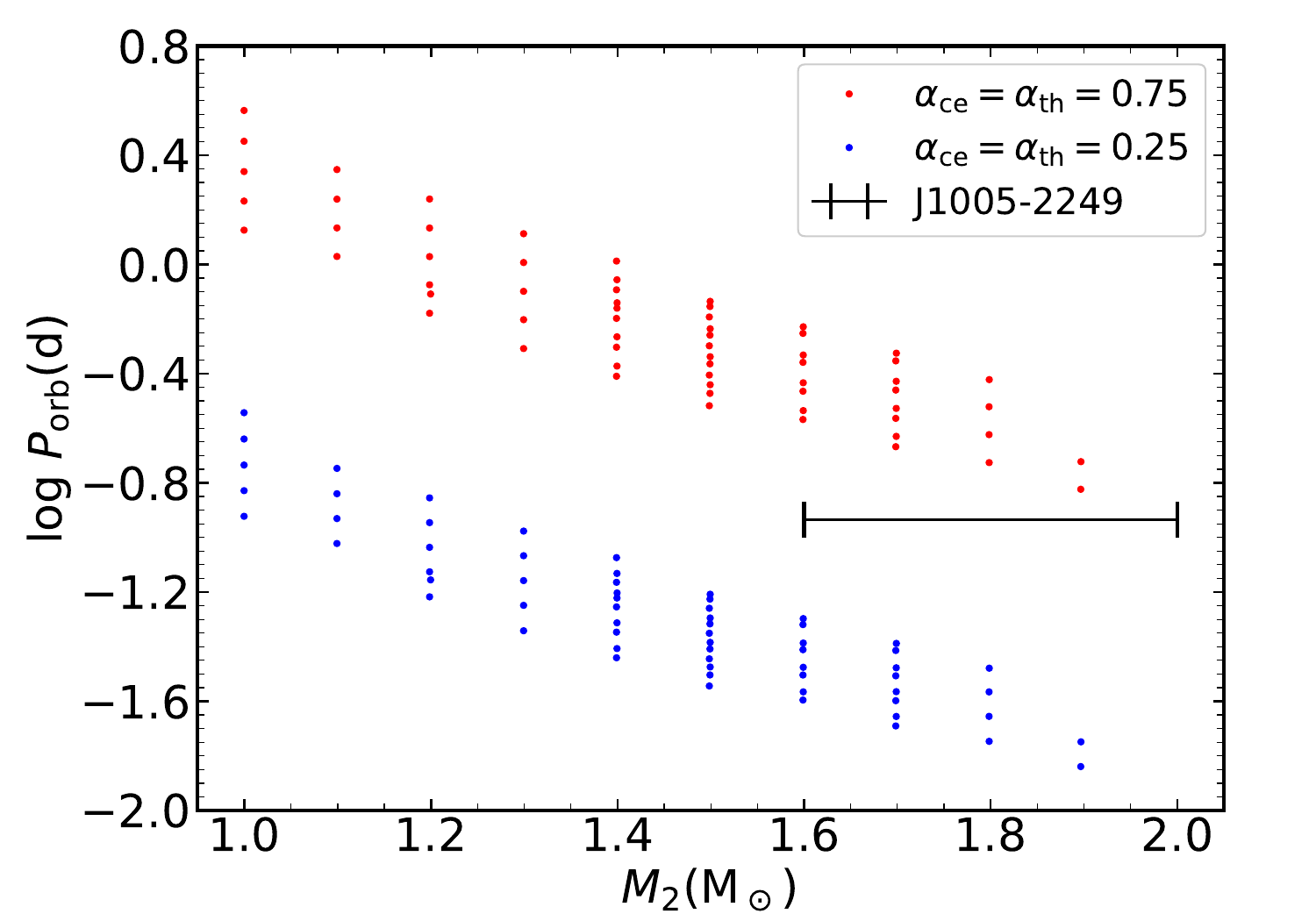}
    \caption{The $M_{\rm 2}$ and $P_{\rm orb}$ for J1005-2249 compared to the model from \texttt{mesa}. The red and blue dots represent the models from \texttt{mesa} with $\alpha_{\rm ce} = \alpha_{\rm th} = 0.75$ and $\alpha_{\rm ce} = \alpha_{\rm th} = 0.25$, respectively. For each $M_{\rm 2}$ with given the $\alpha_{\rm ce}$ and $ \alpha_{\rm th}$, the orbital period is larger for a larger $M_{\rm{WD, 2}}$.}
    \label{fig:m-P-J1005}
\end{figure}

The second WD of J1005-2249 cooling age is $25.5-31.5 {\rm Myr}$, and its orbital period changes small ($\textless$ 1 minute) after the CE ejection\citep{Scherbak2023}. We use its orbital period of 0.11602 days as the binary formation orbital period. In Fig.~\ref{fig:m-P-J1005}, we show $M_{2}$ and orbital period relations with different CE efficiencies and compare them to the J1005-2249. For the case of $\alpha_{\rm ce}=\alpha_{\rm th}=0.75$, the final orbital periods in the simulations are marginally higher than the observational period, which suggests that the CE efficiencies for J1005-2249 should be $\lesssim 0.75$. The results of small CE efficiency with $\alpha_{\rm ce}=\alpha_{\rm th}=0.25$ cannot reproduce the observational period, meaning the CE efficiency of J1005-2249 should be larger than $0.25$. The approximately middle position of J1005-2249 between the results of $\alpha_{\rm ce}=\alpha_{\rm th}=0.75$ and $0.25$ indicates the CE efficiency for J1005-2249 is around $0.5$.

\subsection{WD0957-666}
\begin{figure}
    \centering
    \includegraphics[width=\columnwidth]{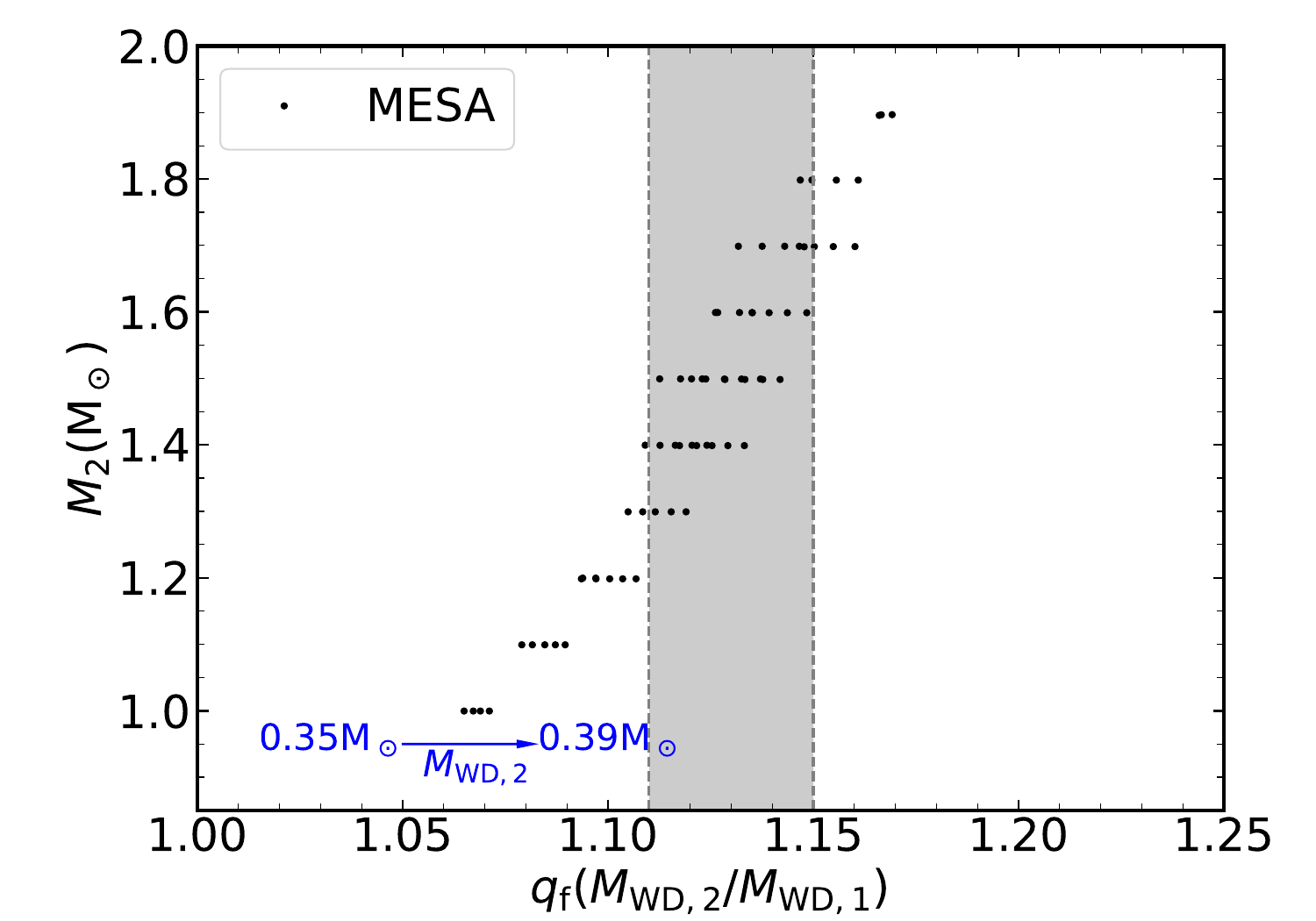}
    \caption{The $q_{\rm f}$ and $M_{\rm 2}$ of WD0957-666 compared to the model from \texttt{mesa}. The gray dots represent the models from \texttt{mesa} with $Z = 0.02$. The gray shade represents $q_{\rm f}$ and the progenitor mass ($M_{\rm 2}$) for the WD0957-666.}
    \label{fig:q-m-WD0957}
\end{figure}

 The mass ratio of WD0957-666 is $q = 1.13\pm 0.02$, derived from its radial velocity amplitudes \citep{Moran1997, Maxted2002b}. The first WD mass of WD0957-666 is $M_{\rm WD,1} = 0.32\pm 0.03 \rm M_{\odot}$, and the second WD mass is $M_{\rm WD,2} = 0.37 \pm 0.02 \rm M_{\odot}$. From Eq.~\ref{6}, we show the $q_{\rm f}$ and $M_{2}$ relation by adopting the $M_{\rm WD,2} = 0.37 \pm 0.02 \rm M_{\odot}$ in Fig.~\ref{fig:q-m-WD0957}. We find that the $M_{2}$ of J1005-2249 range from 1.3 $\rm M_{\odot}$ to 1.8 $\rm M_{\odot}$. 

\begin{figure}
    \centering
    \includegraphics[width=\columnwidth]{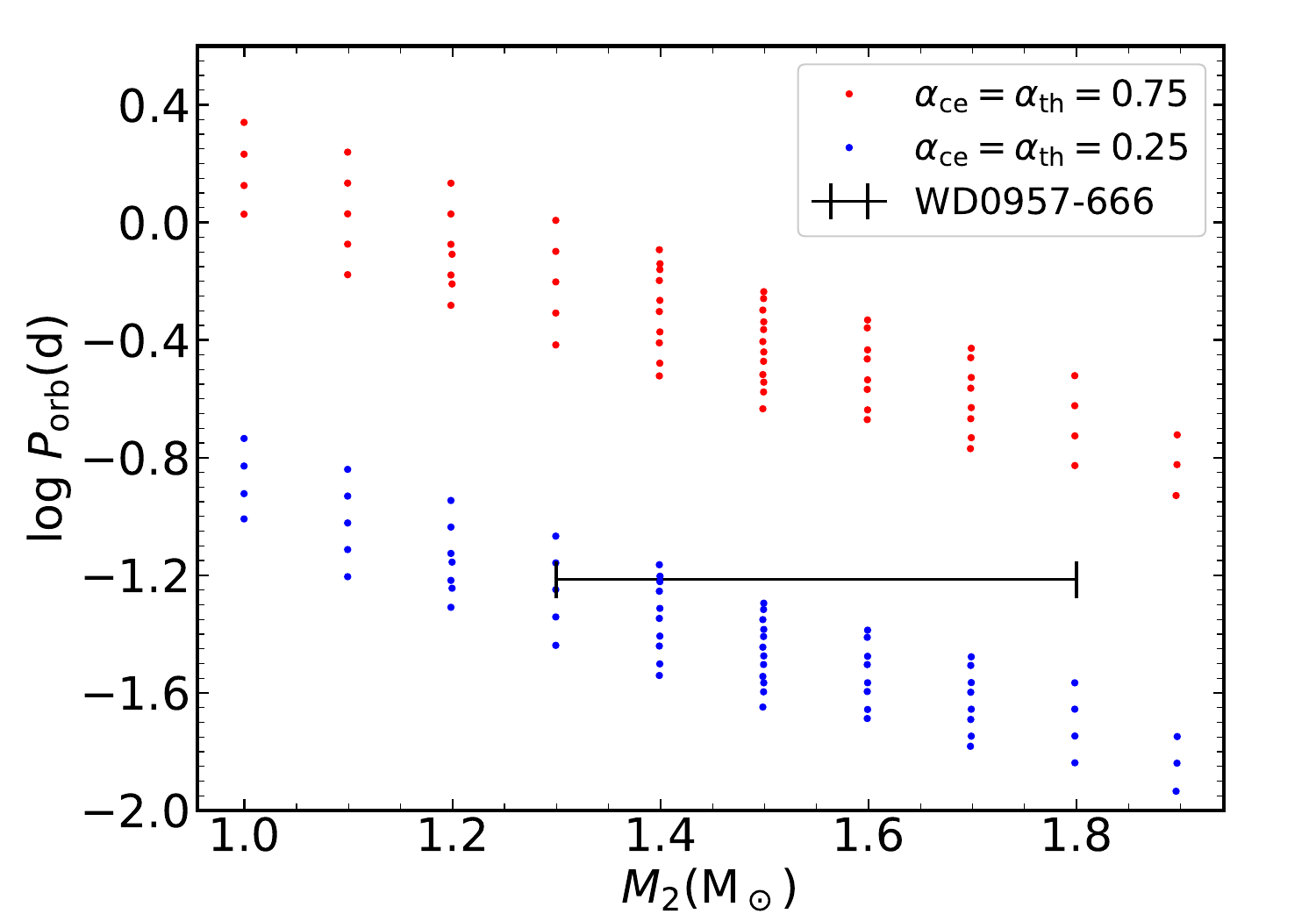}
    \caption{The $M_{\rm 2}$ and $P_{\rm orb}$ of WD0957-666 compared to the model from \texttt{mesa}. The red and blue dots represent the models from \texttt{mesa} with $\alpha_{\rm ce} = \alpha_{\rm th} = 0.75$ and $\alpha_{\rm ce} = \alpha_{\rm th} = 0.25$, respectively.}
    \label{fig:m-P-WD0957}
\end{figure}

The second WD of J1005-2249 cooling age is $2.0-22.2 {\rm Myr}$; its orbital period changes $1-5$ minutes after the CE ejection \citep{Scherbak2023}. So, we also use the observed orbital period 0.06099 day as the binaries formation orbital period. In Fig.~\ref{fig:m-P-WD0957}, we show $M_{2}$ and orbital period relations with different CE efficiencies and compare them to the WD0957-666. We find that the CE efficiencies should be around 0.25 and smaller than 0.75.   
As we see, the second WD masses of J1005-2249 and WD0957-666 are nearly the same. The progenitor mass of the second WD of J1005- 2249 is larger than that of WD0957-666. By comparing the CE efficiencies for the two systems, the CE efficiencies are larger for a larger $M_{2}$. Our results are also consistent with the results from \citet{Nandez2016}.

\section{Summary and discussion}

In this paper, we studied the properties of double He WDs that formed from a particular channel. In this channel, the primary of a binary evolves to RGs with a degenerate core and produces the first He WD via the stable RLOF. Then, the secondary evolves to RGs with a degenerate core and forms another He WD via CE ejection. The main results are as follows.

First, by using a semi-analytic method, the mass ratios of double He WDs are $q_{\rm{f}} \sim 1$, and the second WD is larger than the first WD. To confirm these results, we evolve a grid of MS binaries to form the double He WDs. The outcomes also show that the mass ratios of double He WDs are around 1, and the second WD mass is larger than the first WD in most cases. Moreover, we find our results support the observations well.

Second, if we know the two He WD masses (or a He WD mass and mass ratio), the progenitor mass of the second WD can be derived from Eq.~\ref{6}. Then, with Eq.~\ref{11}, the orbital period can constrain the CE efficiencies.
On the observational side, for the J1005-2249 and WD0957-666, we infer the progenitor masses of the second WD are $1.6-2.0 \rm M_{\odot}$ and $1.3-1.8 \rm M_{\odot}$, respectively. Then, we find the CE efficiencies are around 0.5 for the J1005-2249 and 0.25 for the WD0957-666. The CE efficiencies are larger for a larger progenitor mass of the second WDs. It suggests that the values of CE efficiencies may not be constant (see \citealt{Geh2022, Geh2023}, who found similar results in modeling sdB binaries). 

If the double CO WDs are formed via the RLOF + CE channel, the progenitors of the two WDs also have degenerate core on AGB, so the two WD masses have a similar relation as the Eq.~\ref{6}. Then, the mass ratios can also be used to study the double CO WDs. We confirm that the mass ratio is important in studying the CE ejection for the double WDs. If more double WDs are observed with high precision, the CE efficiencies can be known precisely for the CE occurring on the RGB or AGB.

\section{Acknowledgements}
 
We would like to thank an anonymous referee for very constructive suggestions which improves the manuscript a lot. We also thank Hailiang Chen, Jiao Li, Jiangdan Li for useful comments. This work is supported by the National Key R$\&$D Program of China (grant Nos. 2021YFA1600403, 2021YFA1600400), the National Natural Science Foundation of China (Grant No. 12125303, 12288102, 12090040/3, 12103086, 12273105, 11703081, 11422324, 12073070), the Natural Science Foundation of Yunnan Province (No. 202201BC070003) and the Yunnan Revitalization Talent Support Program–Science $\&$ Technology Champion Project (No. 202305AB350003), the Key Research Program of Frontier Sciences of CAS (No. ZDBS-LY-7005), Yunnan Fundamental Research Projects (grant Nos. 202301AT070314, 202101AU070276, 202101AV070001), and the International Centre of Supernovae, Yunnan Key Laboratory (No. 202302AN360001), High-level talent scientific research start-up fund of Zhoukou Normal University (ZKNUC2021011).
  


%


\bibliographystyle{aasjournal}
\bibliography{zyy_refs}

\appendix
\section{Stellar evolutionary model}
\label{App}
In order to calculate the CE ejection using the semi-analytic method, we need to calculate the binding energy. We then use the \texttt{mesa} (version 10398; \citealt{Paxton2011, Paxton2013, Paxton2015, Paxton2018, Paxton2019}) to compute several stellar models with initial masses of 1.0, 1.26, 1.6, 1.8 $\rm M_{\sun}$ for Population I and Population II. We set the initial hydrogen abundance as $X = 0.76-3Z$ \citep{Pols1998}. The mixing length parameter $\alpha = l/H_{\rm p}$ is adopted to be 2. We implement a step-function overshooting model, assuming that the overshooting region extends by a distance of ${0.25 H_{\rm p}}$ \citep{Pols1998, Hurley2000, Han2002}. Regarding the stellar winds, we apply the prescription from \citet{Reimers1975} for $\eta = 0.25$. Furthermore, the OPAL type II opacity table is adopted \citep{Iglesias1996}. We define He core mass + 0.01$\rm M_{\sun}$ as the boundary to calculate the envelope binding energy \citep{Han1995}. 

\begin{figure}
   \centering
   \includegraphics[width=7.5cm]{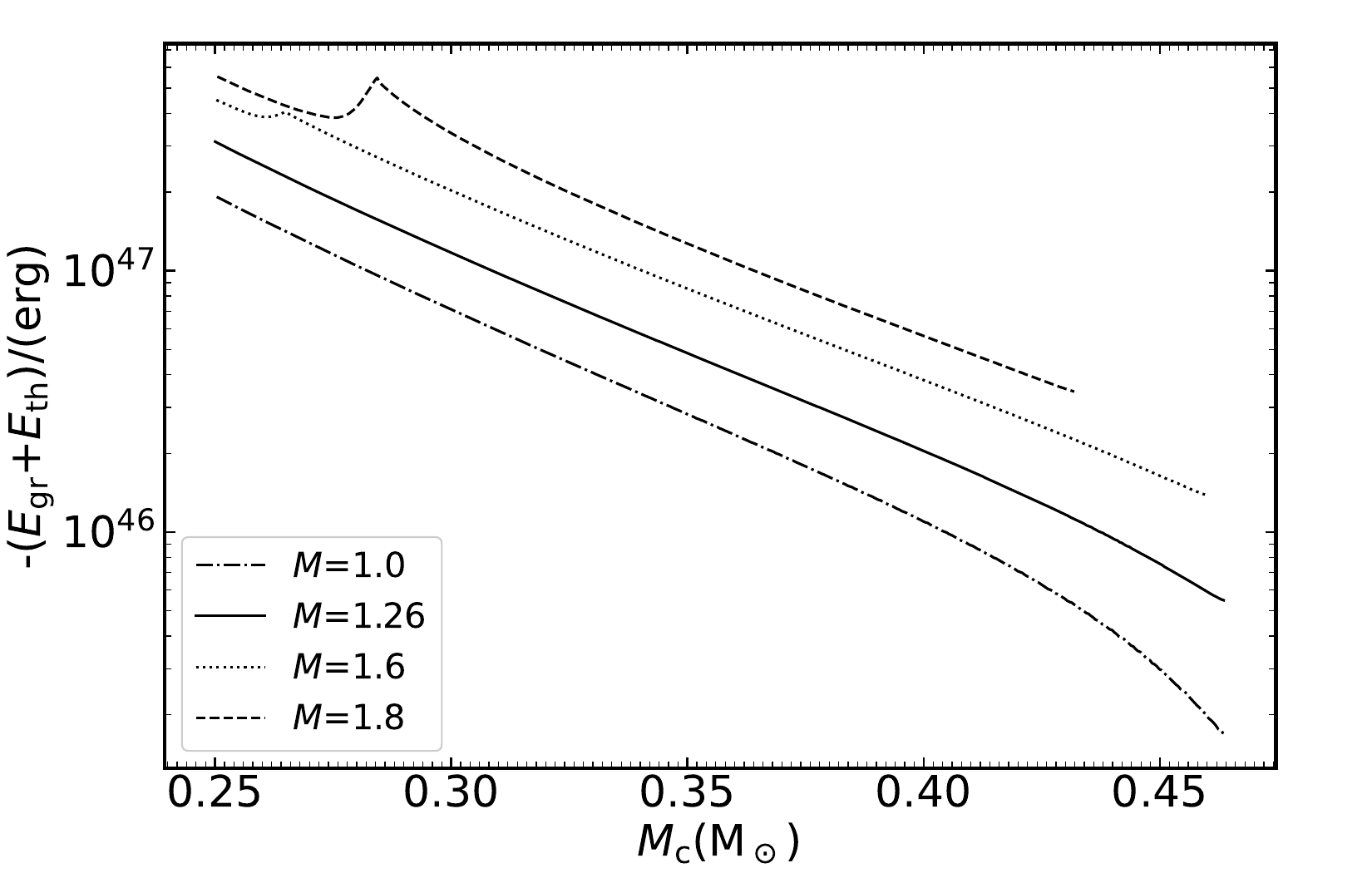}
      \caption{Binding energy as a function of He core mass for four models with typical initial masses: 1.0, 1.26, 1.6, 1.8 $\rm{M_{\sun}}$ at Z = 0.02. In these models, $\alpha_{\rm th} = 1.00$ is adopted.}
         \label{A.1}
\end{figure}

With Eq.~\ref{8} and $\alpha_{\rm th}$, we derive the binding energy for various RGs based on their core mass. We show some models with $\alpha_{\rm th} = 1.00$ for the $Z = 0.02$ in Fig.~\ref{A.1}.

\section{The influence of mass transfer efficiency}
\label{App2}
To show the effect of mass transfer efficiency on our results, we calculated a series of binary models with $\beta = 0.5$ for J1005-2249 and WD0957-666. The initial primary masses range from 1.0 to 1.5 $\rm M_{\odot}$ in steps of $\Delta{M} = 0.1 {\rm M_{\odot}}$, and the secondary masses range from 0.9 to 1.45 $\rm M_{\odot}$ in steps of $\Delta{M} = 0.05 {\rm M_{\odot}}$. The initial orbital periods vary from 3.0 to 75 days, with logarithmic increments of $0.1$.

The main effects of mass transfer efficiency on the results are introduced as follows.
Firstly, the models with $\beta=0.5$ more easily experience unstable mass transfer processes (mass transfer rate larger than $10^{-4}M_\odot\;\rm yr^{-1}$ or the secondary filling its Roche lobe due to the accretion) at the first mass transfer.
In a special case, if the progenitor of the secondary has developed a helium core and the envelope is ejected successfully, the binary will produce a DWD with both WD components having the same birth time \citep{Brown1995b, Justham2011}.
However, the two WD components for J1005-2249 and WD0957-666 show different cooling ages, according to \citet{Bours2014} and \citet{Maxted2002b}. Therefore, the J1005-2249 and WD0957-666 will not likely be produced in this way. 
Secondly, the mass transfer efficiency can affect the binary parameter space that produces the DWDs through the RLOF + CE channel with degenerate cores. For example, the initial progenitor masses of the WDs are 1.0 - 2.0 $\rm M_{\odot}$ for $\beta = 0$ but 0.9 - 1.5 $\rm M_{\odot}$ for $\beta = 0.5$.

\begin{figure}
   \centering   \includegraphics[width=15cm]{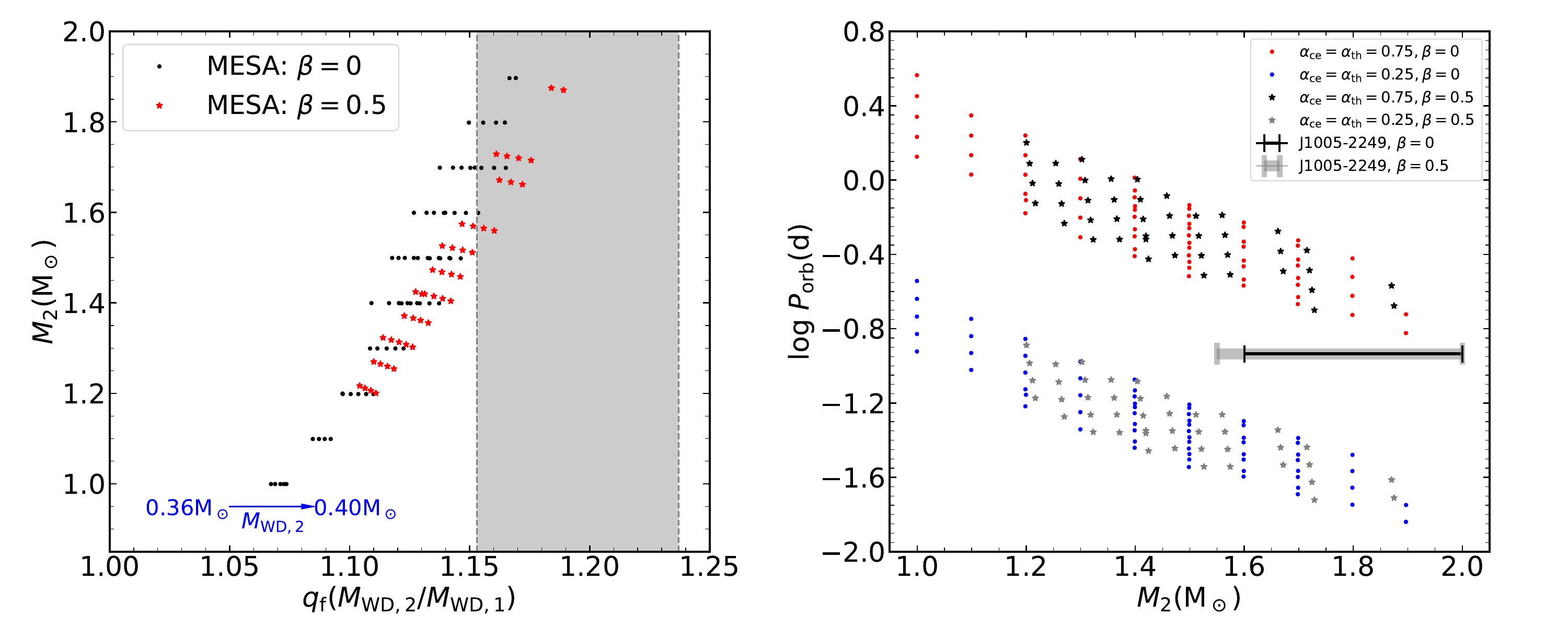}
      \caption{Similar to the Fig.~\ref{fig:q-m-J1005} and Fig.~\ref{fig:m-P-J1005}, but for the results of different mass transfer efficiencies. The cases of $\beta$ = 0.5 and $\beta$ = 0 are shown in stars and dots, respectively. For J1005-2249, the $M_{2}$ is in the range of 1.55 - 2.0 $\rm M_{\odot}$ with $\beta = 0.5$, in the range of 1.6 - 2.0 $\rm M_{\odot}$ with $\beta = 0$.}
         \label{A.2}
\end{figure}

\begin{figure}
   \centering   \includegraphics[width=15cm]{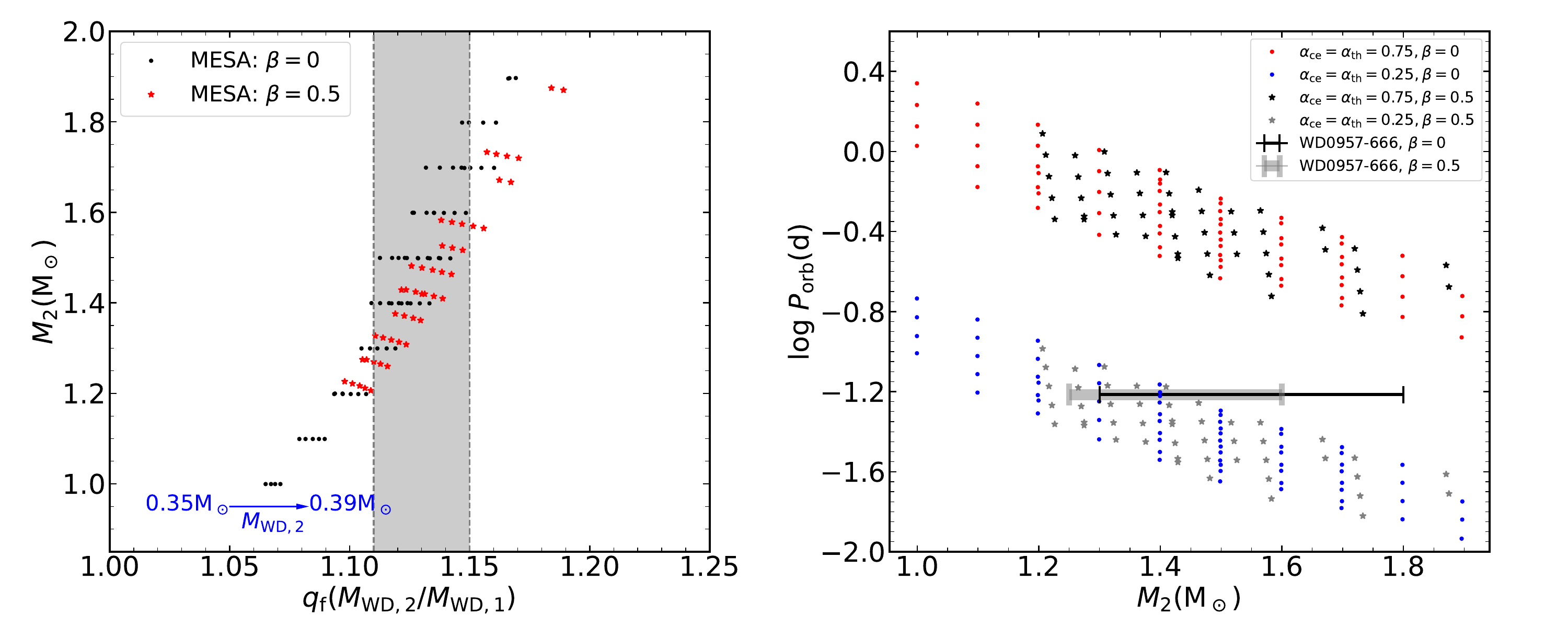}
      \caption{Similar to the Fig.~\ref{fig:q-m-WD0957} and Fig.~\ref{fig:m-P-WD0957}, but for the results of different mass transfer efficiencies. The cases of $\beta$ = 0.5 and $\beta$ = 0 are shown in stars and dots, respectively. For WD0957-666, the $M_{2}$ is in the range of 1.25 - 1.6 $\rm M_{\odot}$ with $\beta = 0.5$, in the range of 1.3 - 1.8 $\rm M_{\odot}$ with $\beta = 0$.}
         \label{A.3}
\end{figure}

In Fig.~\ref{A.2} and Fig.~\ref{A.3}, we compare the results of the models with $\beta=0$ and $0.5$ for J1005-2249 and WD0957-666, respectively. We see that the relation between $M_2$ and $q_{\rm f}$ slightly changes with the mass transfer efficiencies. This is mainly because the range of initial primary mass is different between the cases of $\beta=0$ and $0.5$. For the case of $\beta=0.5$, the primary masses are typically less than 1.6 $\rm M_{\odot}$, and then the produced degenerate cores follow the ${M_\mathrm{c}}$ - $R$ relation during the first mass transfer phase. However, in the case of $\beta=0$, some primary stars have masses larger than $1.6M_\odot$, and the produced cores are partially degenerate. These stars do not strictly follow the ${M_\mathrm{c}}$ - $R$ relation during the first mass transfer phase. As a result, the $M_2-q_{\rm f}$ relation shows a minor change for different mass transfer efficiencies. Nevertheless, the mass transfer efficiency has a limited effect on our main conclusion, i.e., constraining the progenitor mass of the secondary with DWD mass ratio and the CE efficiency.

\end{document}